\documentclass[12pt]{article}

\usepackage{amscd}
\usepackage{cite}

\usepackage{verbatim}

\usepackage{amssymb}

\usepackage{amsmath}

\usepackage{amsthm}

\begin{document}

\begin{center}

\vspace{24pt} { \large \bf A naked singularity stable under scalar field perturbations} \\

\vspace{30pt}

\vspace{30pt}

\vspace{30pt}

{\bf Amruta Sadhu} \footnote{sadhuamruta@students.iiserpune.ac.in}, {\bf Vardarajan
Suneeta}\footnote{suneeta@iiserpune.ac.in}

\vspace{24pt} 
{\em  The Indian Institute of Science Education and Research (IISER),\\
Pune, India - 411021.}

\end{center}
\date{\today}
\bigskip

\begin{center}
{\bf Abstract}
\end{center}
We prove the stability of a spacetime with a naked singularity under scalar field perturbations, where the perturbations are regular at the singularity. This spacetime, found by Janis, Newman and Winicour, and independently by Wyman, is sourced by a massless scalar field and also arises as a certain limit of a class of charged dilatonic solutions in string theory. This stability result opens up specific questions for investigation related to the cosmic censorship conjecture and the mechanism by which it is implemented in nature.

\newpage

\section{Introduction}
\setcounter{equation}{0} There is a vast body of work on the issue of classical stability of black holes under perturbations. As is well-known, the region exterior to the horizon of the Schwarzschild black hole is linearly stable under perturbations by various fields \cite{sbhstable, vish, price, moncrief, wald}. Specifically, there are stability results for perturbations (for e.g., scalar field or gravitational perturbations) that are initially of compact support and obey Dirichlet or Neumann boundary conditions at the horizon and infinity. These stability results have also been shown in Kruskal coordinates, which are valid across the horizon. However, since the perturbations live in the spacetime domain exterior to the black hole horizon, they cannot access the black hole curvature singularity, and it is not the boundary at which boundary conditions are enforced. \footnote{Other classes of perturbations, such as quasinormal modes have ingoing boundary conditions at the horizon, which are motivated by the presence of the curvature singularity in the black hole interior. However, they are not normalizable and are thus different from the perturbations considered in the usual classical stability analysis.}

However, one could now ask similar stability questions for situations in which the perturbations live in a spacetime domain with a curvature singularity, or when the curvature singularity is a natural boundary at which boundary conditions need to be enforced. An example of this is the question of stability of a spacetime containing a naked singularity under various types of perturbations. Such a spacetime would be geodesically incomplete, as can be probed by point particle probes which move along geodesics. Suppose, instead of point particle probes, if fields were used as probes (i.e., the spacetime was perturbed by a field), then would the time evolution of the field be well-defined in a spacetime with a naked singularity? This question has been addressed by Wald \cite{wald1} and later by many authors (see, for example, \cite{ishiwald}, \cite{horowitz}, \cite{ishihosoya}, \cite{blau} and references therein). In \cite{giveon}, the authors explicitly show an example where the time evolution of various fields is well-defined in such a spacetime. If the time evolution of fields is well-defined, then the next question is one of the stability of the spacetime. If the spacetime is perturbed by a field configuration which is initially of compact support, for example, then how does this perturbation evolve? The stability of the negative mass Schwarzschild spacetime (whose curvature singularity now becomes naked, as there is no horizon) under gravitational perturbations was analyzed in \cite{gibbons} --- with boundary conditions demanding regularity of the perturbation at the singularity, this spacetime was shown to be unstable \cite{gleiser}. The Reissner-Nordstrom metric with charge larger than mass (super-extremal) and the Kerr metric with angular momentum larger than mass are spacetimes with naked singularities. They were shown to be classically unstable \cite{gleiser1, gleiser2}. It was shown by Christodoulou \cite{christ} that a naked singularity formed during self-similar collapse of a spherically symmetric massless scalar field was unstable (a similar instability result can be found in \cite{nolan1}). These instability results seem consistent with cosmic censorship and the view that naked singularities are not likely to form from generic initial conditions in collapse situations. One example of a naked singularity stable under gravitational perturbations is the self-similar Vaidya spacetime which is sourced by dust. However this is not a serious challenge to cosmic censorship since the matter sourcing the spacetime is null dust, which forms singularities under collapse in flat spacetime \cite{nolan, nolan2}.

In this paper, we probe a naked singularity spacetime which is a solution to the Einstein equation with spherical symmetry and sourced by a massless scalar field. The spacetime, which we call the Janis-Newman-Winicour (JNW) metric, was found independently by Janis, Newman and Winicour \cite{JNW} and Wyman \cite{wyman} who used different coordinates from each other \footnote{ After this manuscript was sent to the journal, we learnt that this spacetime had been found much earlier, by I Z Fisher, and published (in Russian) in 1948 --- we have included the reference both to the original and the English translation \cite{fisher}.}  --- the two solutions were later shown to be identical, and containing a naked singularity \cite{psjoshi}, \cite{virbhadra}.
This spacetime can also be obtained from a class of charged dilaton black holes found by Gibbons and Maeda which are described by mass, charge and a coupling strength of the dilaton to the Maxwell field \cite{gibbonsmaeda} (the solutions and some features were discussed independently later by Garfinkle, Horowitz and Strominger \cite{garfinkle}). The JNW spacetime is obtained by setting the charge to zero in this class of spacetimes (resulting in a naked singularity), and the dilaton plays the role of the scalar field sourcing the JNW spacetime. More precisely, the JNW spacetime is obtained from the charged dilaton metrics in \cite{garfinkle} by setting $r_{+} = 0$, with $r_{-} \neq 0$ in equations (20-23) of that paper. For other values of $r_{+}$ and $r_{-}$ which correspond to charged dilatonic black holes, evolution of scalar fields in such spacetimes are discussed in \cite{horowitz}. Also, we note that in \cite{ishihosoya}, the scalar field equation in the background of the JNW solution (in the coordinates of Wyman) is stated as an example that is well-posed (when the function space is an appropriate Sobolev space), making the spacetime `wave-regular'. However analysis of its stability has not been carried out in that paper.

When the magnitude of the scalar field is zero, the JNW metric reduces to the Schwarzschild black hole metric, and for any nonzero magnitude of the scalar field, the event horizon of the black hole now degenerates into a naked singularity. We probe this spacetime with a test scalar field and find that the JNW spacetime is \emph{stable} with respect to scalar field perturbations --- this result is irrespective of the mass of the probe field. This is in contrast to studies of perturbations of various naked singularities, like the super-extremal Reissner-Nordstrom metric and Kerr metric with large angular momentum, which are unstable (to gravitational perturbations). Of course, it remains to be seen whether the JNW spacetime continues to be stable with respect to gravitational field perturbations --- this is work in progress. If this were true, it would raise the question of whether the JNW spacetime can form in a realistic situation of collapse of a scalar field for generic initial conditions, and whether this stability result poses a challenge for cosmic censorship.

In the next section, we introduce the JNW metric, and rigorously prove its stability under scalar field perturbations. We do this by first putting appropriate boundary conditions on the perturbation at the naked singularity, and then showing that a perturbation that is initially bounded remains bounded for all time. We conclude with a section discussing the result and the interesting questions that arise from it, particularly relating to whether the cosmic censorship conjecture needs modification, and how cosmic censorship is implemented in nature.

\section{Scalar field in naked singularity}
The Janis-Newman-Winicour (JNW) metric, which is a solution to the Einstein equation with spherical symmetry in the presence of a massless scalar field \cite{JNW}, \cite{wyman} is given by
\begin{eqnarray}
ds^2 = - (1 - b/r)^{\nu} dt^2 + \frac{1}{(1- b/r)^{\nu}} dr^2 + r^2 (1 - b/r)^{1-\nu } d\Omega^2 ;
\label{2.1}
\end{eqnarray}
and $d\Omega^2$ is the standard metric on a unit two-sphere.

The scalar field which is a source for this spacetime is
\begin{eqnarray}
\Phi = \frac{q}{b \sqrt{4\pi}} \ln (1 - b/r) .
\label{2.2}
\end{eqnarray}

The parameter $b$ appearing in the JNW metric is related to the ADM mass of the spacetime $M$ by $b = 2 \sqrt{M^2 + q^2 }$.
$\nu = \frac{2M}{b}$, and when $\nu = 1$ ($q=0$), we recover the Schwarzschild metric as there is no scalar field \footnote{ This metric and the Schwarzschild solution are related by a specific duality which is similar to T-duality \cite{shohreh}.}. As we increase $q$ (the magnitude of the scalar field), $\nu$ decreases from $1$ to $0$. An inspection of the curvature invariants reveals that there is a singularity at $r=b$ when $0 < \nu < 1$ for the Ricci and Weyl scalars as well as the Kretschmann invariant, and we will assume this range for $\nu$ throughout the paper. This singularity is globally naked and the weak energy condition is satisfied for this spacetime \cite{psjoshi}, \cite{virbhadra}. Gravitational lensing near this singularity was studied in \cite{chitre, vnc}, and later in \cite{virbhadra1}. Further properties of point particles in the JNW spacetime and their astrophysical consequences were studied in \cite{psjoshi1} and \cite{psjoshi2}. It is evident from these studies that there are pathologies both in the behaviour of massive particles and light rays in this spacetime, and clear differences between their behaviour near the naked singularity and the behaviour of particles near the Schwarzschild horizon.

The question we want to investigate is that of the stability of this naked singularity spacetime under perturbations. The simplest perturbation we can consider is by another test scalar field of mass $m$, whose magnitude is everywhere smaller than the source scalar field (so that we can use linearized perturbation theory). Let us denote this test scalar field $\chi$. Assuming that the only interaction of this scalar field is through minimal coupling with the background gravitational field, its action is
\begin{eqnarray}
S = \int \sqrt{-g} d^4 x \left [ - \frac{1}{2} g^{\mu \nu} \partial_{\mu} \chi \partial_{\nu} \chi - \frac{1}{2} m^2 \chi^2 \right ].
\label{2.2a}
\end{eqnarray}

The equation of motion of the test scalar field $\chi$ arising from this action is given by the massive Klein-Gordon equation,
\begin{eqnarray}
\frac{1}{\sqrt{-g}} \partial_{\mu}[ \sqrt{-g} g^{\mu \kappa} \partial_{\kappa} \chi ] = m^2 \chi ;
\label{2.3}
\end{eqnarray}
where $g$ denotes the determinant of the metric (\ref{2.1}).\\

Let $f(r) = (1 - b/r)^\nu .$ $$\sqrt{-g} = [ \frac{(1 - b/r)}{f(r)}] r^2 \sin \theta .$$ We now choose the following ansatz for $\chi $:
\begin{eqnarray}
\chi = \frac{\psi}{r} Y_{lm} (\theta, \phi) e^{i \omega t}
\label{2.4}
\end{eqnarray}

$Y_{lm}(\theta, \phi)$ are the spherical harmonics, and $\omega$ can be complex. We will prove in two stages that with reasonable physical boundary conditions, the only solutions are those with $\omega$ real. This implies that there are no solutions growing exponentially in time.

With the ansatz (\ref{2.4}), equation (\ref{2.3}) becomes
\begin{eqnarray}
- \frac{d}{dr} \left [(1- \frac{b}{r}) \frac{d\psi}{dr}\right ] + \left [ \frac{b}{r^3 } + \frac{l(l+1)}{r^2 } + m^2 \frac{(1- \frac{b}{r})}{f}  \right ]\psi = \omega^2 \left (1- \frac{b}{r}\right )^{(1-2\nu)}\psi~.~~~~
\label{2.5}
\end{eqnarray}

Define the coordinate $r_{*}$ by $dr_{*} = dr/(1- \frac{b}{r})$. Note that \emph{this is not the usual tortoise coordinate} that changes the $r-t$ part of the metric into a conformally flat form. This coordinate change is mainly for convenience in analyzing (\ref{2.5}). Then (\ref{2.5}) becomes
\begin{eqnarray}
- \frac{d^2 \psi}{dr_{*}^2} + (1- \frac{b}{r}) \left [ \frac{b}{r^3 } + \frac{l(l+1)}{r^2 } + m^2 \frac{(1- \frac{b}{r})}{f} \right ]\psi = \omega^2 \left (1- \frac{b}{r} \right )^{2(1-\nu)}\psi~.~~~~
\label{2.6}
\end{eqnarray}

$r_{*} = r + b \ln \frac{(r-b)}{b}$, and so, as $b \leq r < \infty$, $-\infty \leq r_{*} \leq \infty $. The range of $r_{*}$ is infinite; however, for the actual tortoise coordinate of this metric, the location of the singularity is at \emph{a finite value} of the tortoise coordinate. Due to the range of the new coordinate $r_{*}$ not being finite, (\ref{2.6}) is a singular Sturm-Liouville differential equation of the form
\begin{eqnarray}
M \psi = \omega^2 w(r) \psi,
\label{2.7}
\end{eqnarray}
where
\begin{eqnarray}
M &=& - \frac{d^2 }{dr_{*}^2} + (1- \frac{b}{r}) \left [ \frac{b}{r^3 } + \frac{l(l+1)}{r^2 } + m^2 \frac{(1- \frac{b}{r})}{f} \right ] ; \nonumber \\w(r) &=& \left (1- \frac{b}{r} \right )^{2(1-\nu)}.
\label{2.8}
\end{eqnarray}

First we would like to prove that $\omega^2$ has to be real. This follows if $M$, acting on the space of functions which are square integrable with respect to the measure $w(r)dr_{*}$ is self-adjoint. For this singular Sturm-Liouville problem, the relevant question is whether there exist a choice of boundary conditions for which $M$ is self-adjoint (or has a self-adjoint extension). \footnote{ In \cite{ishihosoya}, it is mentioned that $M$ acting on an appropriate Sobolev space, is essentially self-adjoint --- however, no further details are given.} Our function space is the space of square integrable functions with respect to the measure $w(r)dr_{*}$, which is the natural choice for the  Sturm-Liouville problem (\ref{2.8}). Therefore, we would like to do a careful analysis of self-adjointness of $M$ on this function space and find the appropriate physical boundary conditions to impose on the perturbation. For this, we follow the standard procedure in literature on ordinary differential equations (see for example, the classic paper of Krall and Zettl \cite{kz}, or the standard book \cite{codlev}) and  classify the endpoints of the interval on which the equation is defined, as regular or singular (of limit point type or limit circle type). The endpoints in this problem are $r_{*} = \pm \infty$. They are not finite, hence singular. For the Sturm-Liouville problem defined on an interval $(-\infty, \infty )$ with $(\omega^2) \in \mathbb{C}$ (complex numbers), the singular endpoint $- \infty$ is of \emph{limit circle} type if all solutions to (\ref{2.8}) are square-integrable with respect to the measure $w(r)dr_{*}$ on $(-\infty, \beta)$ for some $\beta \in (-\infty, \infty )$. This classification is independent of $\omega^2 $ --- if this statement is true for some $(\omega^2) \in \mathbb{C}$, then it is true for all $(\omega^2) \in \mathbb{C}$ (for a proof, see \cite{naimark}). A similar definition holds for the other endpoint. If a singular endpoint is not of limit circle type, it is of \emph{limit point} type. Note that the endpoint $+ \infty $ being of limit point type means there must exist at least one $(\omega^2) \in \mathbb{C}$ for which some solution to (\ref{2.8}) is not square-integrable with respect to the measure $w(r)dr_{*}$ on $(\beta, \infty)$. For endpoints of limit circle type, a careful choice of boundary conditions is needed to ensure self-adjointness of $M$. This analysis of self-adjointness is equivalent to the approach in terms of deficiency indices that is more commonly found in quantum mechanics books \cite{kz}. We will summarize the choice of boundary conditions that lead to self-adjointness, but let us investigate first whether the endpoints in our problem are of limit circle type. \\

(i) Endpoint $r_{*} = -\infty$: \\

As $r_{*} \to - \infty$ (that is, $r \to b$), $ r_{*} \sim b \ln \frac{(r-b)}{b}$. In this limit, (\ref{2.6}) becomes
\begin{eqnarray}
- \frac{d^2 }{dr_{*}^2} \psi + C e^{r_{*}/b} \psi &=& \omega^2 e^{2(1- \nu )r_{*}/b } \psi ~~~~; \nonumber  \\
C &=& \frac{1}{b^2} + \frac{l(l+1)}{b^2}.
\label{2.9}
\end{eqnarray}

If all solutions to this equation are square integrable with respect to the measure $w(r)dr_{*}$ on $(-\infty, \beta)$ for some $\beta \in (-\infty, \infty )$ and for $\omega^2 = 0$, then this will be true for all  $(\omega^2) \in \mathbb{C}$ .
When $\omega^2 = 0$, (\ref{2.9}) reduces to

\begin{eqnarray}
- \frac{d^2 }{dr_{*}^2} \psi + C e^{r_{*}/b} \psi &=& 0.
\label{2.10}
\end{eqnarray}

We change variables to $\sqrt{x}$, where $x = C e^{r_{*}/b}$ and rewrite the above as a modified Bessel equation whose solutions are the modified Bessel functions of order $0$. As $r_{*} \to - \infty$, the general solution is of the usual form $$\psi \sim A_1 ~I_{0}(2 b \sqrt{C} e^{r_{*}/2b} ) ~+ ~A_2 ~K_{0}(2 b \sqrt{C} e^{r_{*}/2b} ).$$ In the limit $r_{*} = -\infty$, what is relevant is the behaviour of the modified Bessel functions for small argument. In this limit, $$I_{0}(2 b \sqrt{C} e^{r_{*}/2b} ) \sim ~const.; ~~~K_{0}(2 b \sqrt{C} e^{r_{*}/2b} ) \sim (const.) r_{*}.$$ Both these linearly independent solutions are square-integrable with respect to the measure $w(r)dr_{*}$ on $(-\infty, \beta)$ for some $\beta \in (-\infty, \infty )$. This is because of the exponential fall-off of the measure function $w(r)$ as a function of $r_{*}$ in the limit $r_{*} \to - \infty$. More precisely, as $r_{*} \to - \infty$,
\begin{eqnarray}
\psi \sim A_1 ( 1 + b C (r -b) +.....) + A_2 ( - \ln \frac{(r-b)}{b} + ....).
\label{2.10a}
\end{eqnarray}
and $$\lim_{a \to -\infty} \int_{a}^{a + \beta} \psi^{2} w(r) dr_{*} = \int_{b}^{b + \epsilon} \psi^{2} (1 - b/r)^{2(1 - \nu) -1} ~dr $$ is finite.
So we conclude that the endpoint $r_{*} = -\infty$ is of limit circle type for all $0 < \nu < 1$ (also, the result is independent of the scalar field mass $m$ --- the $m$-dependent term is negligible near this endpoint). This means that we need to choose our boundary conditions carefully at this endpoint.

For completeness, we would like to add that (\ref{2.10}) is indeed the leading approximation to (\ref{2.6}) when $r \to b$ and $0 < \nu < 1/2$. When $\nu = 1/2$, the leading approximation to (\ref{2.6}) is
\begin{eqnarray}
- \frac{d^2 }{dr_{*}^2} \psi + ( \frac{C - \omega^2}{b} ) e^{r_{*}/b } \psi = 0;
\label{2.11}
\end{eqnarray}
and when $1/2 < \nu < 1$, it is
\begin{eqnarray}
- \frac{d^2 }{dr_{*}^2} \psi - \omega^2  e^{2(1- \nu )r_{*}/b } \psi = 0.
\label{2.12}
\end{eqnarray}

The solutions to (\ref{2.11}) and (\ref{2.12}) are also given in terms of modified Bessel functions of order $0$ --- only the argument of the Bessel functions differ for different ranges of $\nu$.
\\

(ii) Endpoint $r_{*} = + \infty  $:\\

 In the limit $r_{*} \to  \infty$ (that is, $r \to \infty $), (\ref{2.6}) becomes
\begin{eqnarray}
- \frac{d^2 }{dr_{*}^2} \psi + m^2 \psi = \omega^2 \psi .
\label{2.13}
\end{eqnarray}

The general solution in this limit is  $$\psi \sim B_1 e^{i k r_{*} } + B_2 e^{- i k r_{*} };$$ where $ k^2 = (\omega^2 - m^2)$. Clearly there exist choices of $(\omega^2) \in \mathbb{C}$ for which $k$ has a nonzero imaginary part (for e.g., $\omega^2$ real and satisfying $\omega^2 < m^2$ ). In this case, one of the two linearly independent solutions, $e^{i k r_{*} }$ or $e^{- i k  r_{*} }$ will go exponentially to zero as $r_{*} \to  \infty$, and the other would grow exponentially. The measure $w(r)dr_{*} \sim dr_{*}$ in this limit, and thus, only one of these two solutions is square integrable with respect to this measure. We conclude that the endpoint $r_{*} = + \infty  $ is of limit point type.\\

Now, in order to make this Sturm-Liouville problem --- i.e., the operator $M$ defined in (\ref{2.8})--- self-adjoint, we need to place appropriate boundary conditions which are dictated by this classification. We will closely follow the paper of Krall and Zettl \cite{kz} for this (including their notation).\\ Some notation: we denote the endpoint $a:= - \infty$, $d := + \infty$ . Then $a$ is of limit circle type and $d$ is of limit point type. Also denote $$\psi (a) = \lim_{r_{*} \to a^+ } \psi (r_{*}) = \lim_{r_{*} \to -\infty } \psi (r_{*}) .$$ We can define $\psi (d)$ similarly.

We need no boundary conditions at endpoint $d$ since it is of limit point type and one of the two linearly independent solutions near $d$ is excluded anyway as it is not square integrable. To specify the self-adjoint boundary conditions at $a$, we first look for solutions to the equation $M y = 0$, i.e
\begin{eqnarray}
- \frac{d^2 y}{dr_{*}^2} + (1- \frac{b}{r}) \left [ \frac{b}{r^3 } + \frac{l(l+1)}{r^2 } + m^2 \frac{(1- \frac{b}{r})}{f} \right ]y = 0.
\label{2.14}
\end{eqnarray}
Denoting the Wronskian of two functions $f$ and $g$ by $W(f,g) = f g' - g f'$, we need to choose two solutions to the equation (\ref{2.14}), $y_1$ and $y_2$ so that for the entire range of $r_{*}$, $$W(y_1 , y_2 ) = 1. $$ We note that standard results (for e.g, Abel's identity) imply that for a homogeneous second order ODE of the type (\ref{2.14}), the Wronskian of two solutions remains constant in the entire range of $r_{*}$. Therefore, we can look at (\ref{2.14}) in the limit $r_{*} \to -\infty$, and identify two linearly independent solutions $y_1 $ and $y_2 $ with $W(y_1 , y_2 ) = 1 $ in this limit. But the approximate equation in this limit is precisely (\ref{2.10}). Therefore, two linearly independent solutions (in this limit) are $y_1 = A_1 ~I_{0}(2 b \sqrt{C} e^{r_{*}/2b} )$ and $y_2 = A_2 ~K_{0}(2 b \sqrt{C} e^{r_{*}/2b} )$. Recall that the Wronskian of the two modified Bessel functions $I_0 (z)$ and $K_0 (z)$ \emph{with respect to the variable $z$} is $(const.)/z$. Since $z = 2 b \sqrt{C} e^{r_{*}/2b}$, the Wronskian of this pair of modified Bessel functions with respect to the variable $r_{*}$ is a constant. By choosing $A_1 $ and $A_2 $ appropriately, we can find $y_1 $ and $y_2 $ so that their Wronskian is unity. Thus, we have shown that there exist two solutions to (\ref{2.14}), $y_1$ and $y_2$ with $W(y_1 , y_2 ) = 1. $

We now return to the problem of finding self-adjoint boundary conditions at the endpoint $a = -\infty$. The correct boundary conditions are given by
\begin{eqnarray}
& &c_1 W(\psi, y_1 )(a) + c_2 W (\psi, y_2)(a)  \nonumber \\
& &= c_1 \lim_{r_{*} \to a^+ } [\psi y_{1}' - y_{1} \psi'] + c_2 \lim_{r_{*} \to a^+ } [\psi y_{2}' - y_{2} \psi'] =  0.~~~~~
\label{2.15}
\end{eqnarray}

Any constants $c_1 $ and $c_2 $ are allowed, provided at least one of them is nonzero. Our notation in the previous equation was that $$W(\psi, y_1 )(a) = \lim_{r_{*} \to a^+ } W(\psi, y_1 );~~~ W(\psi, y_2 )(a) = \lim_{r_{*} \to a^+ } W(\psi, y_2 ).$$ As $r_{*} \to a^+ $, $y_1 \to const.$ and $y_{1}' \to 0$ (prime denotes derivative with respect to $r_{*}$). As $r_{*} \to a^+ $, $y_2 \to - \infty$ and $y_{2}' \to const.$ Now we know from the computations done at the beginning of this section that as $r_{*} \to a^+ $, $\psi$ is in general, some linear combination of the two modified Bessel functions whose argument depends on the value of the parameter $\nu$. For example, when $ 0 < \nu < 1/2 $, $$\psi \sim A_1 ~I_{0}(2 b \sqrt{C} e^{r_{*}/2b} ) ~+ ~A_2 ~K_{0}(2 b \sqrt{C} e^{r_{*}/2b} ).$$  In this limit, the Bessel function $K_{0}$ is singular, whereas $I_{0}$ goes to a constant. Clearly, from the general form of $\psi$ and the behaviour of $y_1$ and $y_2$ as $r_{*} \to a^+$, there exist some choices of $c_1$ and $c_2$ for which the resulting boundary conditions (\ref{2.15}) imply that  $\psi$ is proportional to the Bessel function $I_0$ (i.e., contains no term proportional to $K_0$). Such boundary conditions would result in a regular $\psi$ since $K_0$ is not regular in this limit. For example, consider $c_1 = 1$ and $c_2 = 0$. We then obtain the boundary condition $$\psi y_{1}'(a) - \psi' y_{1}(a) = 0.$$ Studying the behaviour of solutions as $r_{*} \to a^+$,, we see that this implies $\psi'(a) = 0$. This boundary condition picks the modified Bessel function $I_{0}$ (whose derivative with respect to $r_{*}$ is zero at $a$) and excludes $K_0 $ (whose derivative with respect to $r_{*}$ is nonzero at $a$). It is also easy to see that if $c_2 \neq 0$, then $\psi$ will have a component that is proportional to $K_0$ and will thus not be regular at $a$.

We now argue that from physical considerations, we should choose the boundary condition resulting in a $\psi$ that is proportional to $I_0$ (with no contribution from the $K_0 $ piece). We would like this scalar field to be a \emph{perturbation}, which means its back-reaction can be neglected. Recall that the energy momentum tensor of a scalar field $\chi$ of mass $m$ is
\begin{eqnarray}
T_{\mu \nu} = \partial_{\mu} \chi \partial_{\nu} \chi - \frac{1}{2} g_{\mu \nu} [ g^{\alpha \beta} \partial_{\alpha} \chi \partial_{\beta} \chi + m^2 \chi^2].
\label{2.15a}
\end{eqnarray}

Consider the trace of the energy momentum tensor $T = g^{\mu \nu}T_{\mu \nu}$ --- in the background four dimensional spacetime, this is:
\begin{eqnarray}
T &=& g^{\mu \nu} T_{\mu \nu} =- g^{\mu \nu} \partial_{\mu} \chi \partial_{\nu} \chi - 2 m^2 \chi^2 ; \nonumber \\
&=& g^{tt} (\partial_t \chi)^2  + g^{rr} (\partial_r \chi)^2  + g^{\theta \theta} (\partial_{\theta} \chi)^2 + g^{\phi \phi} (\partial_{\phi} \chi)^2 .
\label{2.15b}
\end{eqnarray}

Recall that $\chi$, given in terms of $\psi$ by the ansatz (\ref{2.4}) is in a complex representation  --- in this case, in (\ref{2.15b}), we need to replace $(\partial_t \chi)^2$, for example, by $ (\partial_t \chi) (\partial_t \bar \chi)$ where $\bar \chi $ is the complex conjugate of $\chi $. Let us now choose a non-regular boundary condition so that  $\psi$ contains a component proportional to $K_0$ as $r \to b$.
As $r \to b$, with the non-regular boundary condition, $$\psi \sim D \ln \frac{(r-b)}{b} ,$$ where $D$ is a constant.
So $$\chi \sim (\frac{D}{b}) \ln \frac{(r-b)}{b} Y_{lm} (\theta, \phi) e^{i \omega t}$$ as $r \to b$. Then each term in the expression for $T$ in (\ref{2.15b}) diverges as $r \to b$. For example, as $r \to b$, $$ g^{tt} (\partial_t \chi) (\partial_t \bar \chi) \sim \left [ \frac{b}{(r- b)} \right ]^{\nu} (\omega^2 ) (\frac{D}{b})^2 \left [ \ln \frac{(r-b)}{b} \right ]^2 Y_{lm}^2 .$$ Similarly, $$ g^{rr} (\partial_r \chi) (\partial_r \bar \chi) \sim \left [ \frac{(r-b)}{b} \right ]^{\nu} (\frac{D}{b})^2 \left [ \frac{1}{(r-b)} \right ]^2 Y_{lm}^2 .$$ Therefore, clearly $T$ diverges in this limit.  It can be checked that the last two terms in (\ref{2.15b}) also diverge. It may be noted that the trace of the energy momentum tensor of the source scalar field also diverges at $r=b$ (due to which there is a curvature singularity at $r=b$). However, upon choosing the non-regular boundary condition, the probe field diverges \emph{at the same rate} as the source as $r \to b$, and considering this boundary condition would lead to a back-reaction that cannot be neglected. Choosing this non-regular boundary condition implies that this scalar field can no longer be considered a `perturbation'. Thus the only physical boundary condition is that with $c_1$ nonzero and $c_2 = 0$ in (\ref{2.15}), and without loss of generality, we can take $c_1 = 1$. With the choice of regular boundary condition, $T$ does not diverge at the same rate as that of the source --- and the divergence is only due to the components of the inverse metric in the trace.  `Regularity boundary conditions' have been imposed before in studies of naked singularities for example, in \cite{giveon}.

 We close with a comment on the various details of our computation. To make our analysis easier, we studied (\ref{2.6}) which was obtained from (\ref{2.5}) by a coordinate change from $r$ to $r_{*}$ and writing the differential equation in Sturm-Liouville form. There is a natural measure associated with such a problem, $w(r)$, and this determines our function space (in this case, all square integrable functions with respect to this measure). The natural question that arises is, what would happen to the main features of the analysis, if we had done a different coordinate change, say from $r$ to the genuine tortoise coordinate (which would change the measure). It can be seen that the main features of our analysis --- an endpoint of limit circle type, and one unambiguous self-adjoint boundary condition corresponding to regularity of the perturbation all remain in the analysis of the resulting Sturm-Liouville problem.

Our conclusion is that, with this physically motivated boundary condition, the Sturm-Liouville problem  (\ref{2.6}) defined on the space of square integrable functions with respect to the measure $w(r) dr_{*}$ is self-adjoint, and therefore its spectrum (given by values of $\omega^2$) is real. However, we will not be guaranteed stability unless the self adjoint operator is positive, i.e, $\omega^2 \ge 0$, so that $\omega$ is real. We will now attempt to show this.\\

{\bf Proof that for  normalizable solutions, $\omega$ is real: }
\vskip 0.5cm
We have already shown that $\omega^2 $ is real.
Let us assume that $\omega^2 \leq 0$, so that $\omega$ is imaginary. Denote $\Omega^2 = - \omega^2 \geq 0$. Let $$ V = \left ((1- \frac{b}{r}) \left [ \frac{b}{r^3 } + \frac{l(l+1)}{r^2 } + m^2 \frac{(1- \frac{b}{r})}{f} \right ] +  \Omega^2 \left (1- \frac{b}{r} \right )^{2(1-\nu)} \right ).$$ Then we can rewrite (\ref{2.6}) as
\begin{eqnarray}
& & - \frac{d^2 \psi}{dr_{*}^2} + V \psi = 0
\label{2.16}
\end{eqnarray}

We have assumed that $\Omega^2 \geq 0$. So $V \geq 0$. With this choice of potential $V$, consider the Schr\"odinger equation
\begin{eqnarray}
- \frac{d^2 \psi}{dr_{*}^2} + V \psi = E \psi .
\label{2.17}
\end{eqnarray}

Then the equation (\ref{2.16}) corresponds to the zero eigenvalue equation ($E=0$) in (\ref{2.17}). Solutions to (\ref{2.17}) with $E \neq 0$ are not connected to our problem, but for the moment, let us consider the eigenvalue problem for the Schr\"odinger equation (\ref{2.17}). Since $V \geq 0$, this implies that for normalizable eigenfunctions to (\ref{2.17}), $E$ cannot be negative. To see this, we multiply (\ref{2.17}) by $\bar \psi$ (complex conjugate of $\psi$) and integrate over the entire range of $r_{*}$. Then after integration by parts, we see that
\begin{eqnarray}
 \left. - \bar \psi \frac{d\psi }{dr_{*}} \right | _{-\infty}^{+\infty} + \int \left |\frac{d\psi }{dr_{*}}\right |^2 dr_{*} + \int V |\psi |^2 dr_{*} = E \int |\psi |^2 dr_{*} .
 \label{2.18}
 \end{eqnarray}
Now we impose the same boundary conditions on $\psi$ that led to self-adjointness for operator $M$. This ensures that the boundary term in (\ref{2.18}) goes to zero. Then it is clear that the lefthand side of (\ref{2.18}) is nonnegative due to $V$ being nonnegative. Thus the right hand side cannot be negative, and $E \geq 0$. So if $E=0$ is an eigenvalue of (\ref{2.18}), it is the ground state eigenvalue. We are interested in the ground state eigenfunction which we denote $\psi_0$ --- this solves (\ref{2.16}) and we would like to know if such a normalizable eigenfunction exists. Recall that the ground state eigenfunction for a Schr\"odinger problem has no nodes (zeroes)in the interval $(- \infty, \infty)$ (for a proof, see \cite{courant}). Since $\psi_0$ is continuous, we can therefore assume without loss of generality that $\psi_0 > 0$ in $(- \infty, \infty)$ (and if it approaches zero, it only does so asymptotically). Consider the equation obeyed by $\psi_0$:
\begin{eqnarray}
 - \frac{d^2 \psi_0 }{dr_{*}^2} + V \psi_0 = 0 .
\label{2.19}
\end{eqnarray}
Then
$$\frac{d^2 \psi_0 }{dr_{*}^2} = V \psi_0  \geq 0, $$ since we can assume $\psi_0 > 0$, and $V \geq 0$. Therefore $\frac{d\psi_0 }{dr_{*}}$ is an increasing function of $r_{*}$. We already know the behaviour of $\psi_0$ as $r_{*} \to - \infty$ from the discussion on this endpoint being of limit circle type. For the choice of boundary conditions made, $\psi_0$ is proportional to a modified Bessel function $I_0$ and as can be checked from the properties of these Bessel functions, $\frac{d\psi_0 }{dr_{*}} > 0$ for $r_{*}$ very negative. Therefore, $\frac{d\psi_0 }{dr_{*}}$ remains positive. This implies $\psi_0$ is an increasing function of $r_{*}$, and furthermore, the modified Bessel function $I_0 \to 1$ as $r_{*} \to - \infty $. It is therefore impossible for $\psi_0$ to go to zero asymptotically as $r \to \infty$ as required by the boundary conditions --- to do so, it must decrease somewhere, but its slope is never negative. Thus we conclude that there is no normalizable solution to the equation (\ref{2.19}). Our argument relied on the positivity of $V$, which in turn is due to assuming that $\omega^2 \leq 0$. The conclusion therefore is that there are no normalizable solutions to (\ref{2.6}) for $\omega$ zero or pure imaginary.

These arguments show that with the appropriate boundary conditions, $M$ is a positive self-adjoint operator acting on the Hilbert space of square integrable functions with respect to the measure $w(r) dr_{*}$ which we denote by $\mathcal{H}$. Rewriting the Klein-Gordon equation (\ref{2.3}) as
\begin{eqnarray}
\frac{\partial^2 }{\partial t^2 } \psi = - \frac{1}{w(r)} M \psi ;
\label{2.20}
\end{eqnarray}
we start with initial data of compact support, denoted by $\psi_{t=0}$ and $(d_t \psi)_{t=0}$, and investigate if the solutions to (\ref{2.20}) that evolve from this initial data lie in $\mathcal{H}$. As discussed by Wald \cite{wald1}, since $M$ is self-adjoint and positive, the answer is in the affirmative --- further, the solutions evolving from initial data of compact support are defined everywhere, smooth, and bounded (in $t$) in the spacetime. Of course, we have obtained this result by choosing a particular boundary condition at the naked singularity. Horowitz and Marolf \cite{horowitz} have raised the issue of different choices of inequivalent boundary conditions leading to different evolution starting from a set of initial data. This is similar to quantum mechanical problems in which the Schroedinger operator is symmetric, but not self-adjoint. In such problems, there may be an infinite number of self-adjoint extensions possible (i.e., inequivalent boundary conditions) each of which corresponds to a different spectrum. The choice of self-adjoint extension (boundary conditions) is dictated by the physics of the problem. In our case, we have done the same --- only one unambiguous choice of boundary conditions implies regularity of the solution (so it remains a `perturbation' and back-reaction effects can be ignored --- otherwise the perturbative analysis in terms of the Klein-Gordon equation on the JNW background would break down). We have shown that for this choice of boundary condition, $M$ is self-adjoint and positive. For other choices of self-adjoint boundary conditions involving the solution which is not regular at the naked singularity, the statements we have made on positivity of $M$, particularly after (\ref{2.18}) are not valid. We do not know if those boundary conditions lead to positive self-adjoint extensions. Our choice of function space was the traditional space of square integrable functions with respect to a measure. For this very example, it has been mentioned in \cite{ishihosoya} that if the function space is an appropriate Sobolev space, $M$ is essentially self-adjoint --- and therefore there is a unique choice of boundary conditions at both endpoints.

This completes the main arguments of this section, and our conclusion is that the JNW spacetime with a naked singularity is classically stable under perturbations which are initially of compact support. Our result is for perturbations due to a massless or massive scalar field which is \emph{not} the source scalar field. This analysis would not be valid for a perturbation of the source scalar field. To deal with the source perturbations consistently, we need to consider the perturbed scalar field equation and the perturbed Einstein equation together and solve for the resulting gravitational perturbations as well. This is work in progress, but we close this section with some comments on this computation. Let the perturbed source field be $\Phi + \delta \Phi $ and resulting metric $g_{\mu \nu} + h_{\mu \nu}$ ($g_{\mu \nu}$ is the JNW metric). Then the equation obeyed by $\delta \Phi$ in linearized perturbation theory is not (\ref{2.3}), but a Klein-Gordon equation with a source (the source is related to  $h_{\mu \nu}$). Denoting the covariant derivative by $\nabla$ and trace of the metric perturbation by $h$, the resulting equation is
\begin{eqnarray}
\nabla_{\mu} \left [ g^{\mu \nu} \partial_{\nu} (\delta \Phi) \right ] = \nabla_{\mu} \left [ h^{\mu \nu} \partial_{\nu} \Phi - \frac{h}{2} g^{\mu \nu} \partial_{\nu} \Phi \right ]
\label{2.21}
\end{eqnarray}

The solution $\delta \Phi $ for this nonhomogeneous PDE is formally written using the Green's function for the problem as well as the solution to the homogeneous equation, since this can be added to any solution to (\ref{2.21}). We have already studied the homogeneous solution (with source terms set to zero) in this paper. To obtain the full solution to (\ref{2.21}) we need to solve the linearized Einstein equation as well (and decouple the equations for the metric and scalar perturbation) . However, let us write the trace of the energy momentum tensor of the perturbed scalar field (in linearized perturbation theory). This is $T_{p} = g^{\mu \nu} \left [ T_{\mu \nu} + \delta T_{\mu \nu} \right ]$. Here, $T_{\mu \nu}$ and $g_{\mu \nu}$ refer to the energy momentum tensor and metric of the background. From the expression for the energy momentum tensor for a scalar field given in (\ref{2.15a}), it can be easily verified that the correction to the trace of the background energy momentum tensor is
\begin{eqnarray}
g^{\mu \nu} \delta T_{\mu \nu} = - 2 \partial_{\mu} (\delta \Phi) g^{\mu \nu} (\partial_{\nu} \Phi ) - \frac{1}{2} h (\partial_{\mu} \Phi ) g^{\mu \nu} (\partial_{\nu} \Phi ) - 2 ( \partial_{\mu} \Phi ) h^{\mu \nu} ( \partial_{\nu} \Phi ).~~~~~
\label{2.22}
\end{eqnarray}

The correction is linear in the scalar and metric perturbations. If the scalar and metric perturbations obey `regularity boundary conditions' and can be described as a Frobenius series around $r=b$, then we can compare $g^{\mu \nu} \delta T_{\mu \nu}$ to the background $T$ \footnote{ With different `ingoing' boundary conditions at the singularity which corresponds to the perturbation not being regular, this spacetime has been shown to be unstable under radial perturbations \cite{bronnikov}.}. Since the background field $\Phi$ diverges logarithmically as $r \to b$, $T$ diverges as some negative power of $(r-b)$. If the scalar and metric perturbations were regular, the correction term $g^{\mu \nu} \delta T_{\mu \nu} $ would either diverge at a slower rate than $T$ or be finite. Thus, back-reaction could be neglected in that case. It is promising that we indeed have a homogeneous solution to the equation (\ref{2.21}) with source terms set to zero which can be written as a Frobenius series around $r=b$. We hope to do a complete analysis of both (\ref{2.21}) and the perturbed Einstein equation for the scalar and metric perturbations to investigate if this feature persists in the solutions to the equations with sources.

\section{Discussion}

In this paper, we have shown that the JNW spacetime, which contains a naked singularity, is stable under scalar field perturbations which are regular at the singularity. This is a very surprising result, as we would have expected naked singularities to be perturbatively unstable. Furthermore, the JNW spacetime is sourced by `realistic' physical matter, a scalar field (unlike naked singularities in fluid models or dust, where the matter model breaks down at small scales). So the obvious generalization would be to consider gravitational perturbations of this spacetime and comment on its stability.
If this spacetime turns out to be stable under gravitational perturbations as well, this would be indicative of the spacetime forming as a result of collapse for generic initial conditions. The question then would be if the cosmic censorship conjecture needs to be modified --- it has been suggested, for example, that the physically relevant condition for a spacetime is not global hyperbolicity, but the well-posedness of all the probe field equations in the spacetime \cite{clarke}.

Another point of view could be that this naked singularity spacetime may have an instability in quantum gravity, even if it is stable classically. Thus, perhaps in those examples where naked singularities are classically stable, quantum gravity effects implement cosmic censorship. This surmise can be tested for this example --- the question of stability of a spacetime in quantum gravity can be examined in the Euclidean path integral formulation \cite{GPY} (also see \cite{vs} and references therein).
Thus, this stability result for the JNW spacetime opens up many tractable questions which will hopefully lead to a precise geometric/analytic criterion for stability of a naked singularity.

\end{document}